\documentclass[english]{article}
\usepackage{babel, mathrsfs, bbm, amsmath,amssymb}
\usepackage{bm}
\usepackage[T1]{fontenc}
\usepackage[latin1]{inputenc}

\title{Parametrizing Fluids in Canonical Quantum Gravity}

\author{
Giovanni Montani\footnote{montani@icra.it} (1 2 3), Simone Zonetti\footnote{simone.zonetti@icra.it} (1)\\
{\small (1) Dipartimento di Fisica, Universit\`a degli Studi di Roma "La Sapienza",} \\
{\small (2) ENEA C.R. Frascati (Dipartimento F.P.N.), (3) ICRANet C.C. Pescara.}
}

\date{February 2008}

\begin{document}

\maketitle

\begin{abstract}
The problem of time is an unsolved issue of canonical General Relativity. A possible solution is the Brown-Kucha\v r mechanism which couples matter to the gravitational field and recovers a physical, i.e. non vanishing, observable Hamiltonian functional by manipulating the set of constraints. Two cases are analyzed. A generalized scalar fluid model provides an evolutionary picture, but only in a singular case. The Schutz' model provides an interesting singularity free result: the entropy per baryon enters the definition of the physical Hamiltonian. Moreover in the co-moving frame one is able to identify the time variable $\tau$ with the logarithm of entropy.
\end{abstract}

\section{The Kucha\v r-Brown mechanism}
The problem of time is a well known issue in the context of the canonical approaches to Quantum Gravity, where the Hamiltonian function is vanishing.
A possible solution is to couple matter to the gravitational field, and use its properties to recover an evolutionary picture through the Brown-Kucha\v r mechanism (see \cite{kucharbrown:1995}). One considers for example a generic scalar field Lagrangian in the form $
\boldsymbol{L}_F = \boldsymbol{L}_F (-\phi_{,\mu} \phi^{,\mu}) = \boldsymbol{L}_F (\Upsilon)$, from which the calculation of the conjugate momentum is straightforward:
\begin{equation}\nonumber \label{gmomentum}
\pi = \frac{\delta \mathcal{S}_F}{\delta \dot \phi } = - 2 \sqrt{q}( \phi_n ) \frac{\delta \boldsymbol{L}_F}{\delta \Upsilon},
\end{equation}
where $\phi_n  = n^\mu \nabla_\mu \phi$; this can be seen as an equation for $\Upsilon$,
which will have some solution $\Upsilon = \tilde{\Upsilon}(\pi, V)$.
Using the expression $\Upsilon = (\phi_n )^2 -\phi_{,a} \phi^{,a} = (\phi_n )^2 - V$ one also obtains $\phi _n = \chi (\pi, V)$. Using $\tilde{\Upsilon}$ and $\chi$ one can perform the Legendre transformation and move to the canonical formalism, so that the Hamiltonian has the form $H_F = \int d^3x \left(H^F_a N^a + H^F N\right)$, with suitable $H^F$ and $H^F_a$ functionals. Here the field $\phi$ appears only through its spatial gradients.
By adding the Einstein-Hilbert action, the total Hamiltonian density will be simply the sum of the two uncoupled densities.
At this point one can square the super-momentum, and express $V$ as a function of the gravitational variables and $\pi$.
Making the substitution \emph{the super-Hamiltonian will contain no fluids variables but $\pi$}. This allows one to solve it too, as an equation for $\pi$, so that the equivalent constraint will read:
\begin{equation}\label{nsH}
\pi - h(H^G, H^G_a) = 0.
\end{equation}
This is the starting point to the construction of a physical Hamiltonian, since one can see in \eqref{nsH} the form of a Schr\" odinger equation.
It will simply consist in $\boldsymbol{H}_{phys} = \int d^3 x h(x)$, and it will need to fulfil some conditions:
\begin{itemize}
\item Independence from the field whose conjugate momentum is $\pi$.
\item Invariance under the 3-diff group. This is manifest if $h$ is a scalar density of weight one.
\item Invariance under the super-Hamiltonian constraint. In the case of multiple solutions this condition may be used to eliminate the unphysical ones.
\end{itemize}
Once these properties are checked, one is able to write the evolution equation for observables\footnote{Here one assumes that observables exist, and have the standard property of vanishing Poisson brackets with the whole set of constraints} as the action of $\boldsymbol{H}_{phys}$ via the Poisson brackets:$-\frac{d \mathcal{O}(t)}{dt}= \{\boldsymbol{H}_{phys}, \mathcal{O}(t)\}$.

\section{Generalized scalar field fluid} 
One considers a special case of the scalar field Lagrangian presented in the previous section, where the dependence on $\Upsilon$ is given by $\mathcal{S}_F = \int dt d^3 x \sqrt{q} N \Upsilon^\gamma$, where $\gamma$ is real valued.
The resulting equation of state for the fluid is $p = \frac{\rho}{2\gamma -1}$, where $\rho$ is the energy density and $p$ is the pressure. This gives a constraint on the values taken by $\gamma$, since $p$ and $\rho$ need to be positive: $\gamma > 1/2$.
Moreover the square of the velocity of sound in the medium is $S^2 = \frac{c^2}{2 \gamma - 1}$, so it gives a more strict constraint, fixing $\gamma \geq 1$.\\
Using the definition of the momentum conjugate to $\phi$, one obtains the equation for $ \phi_n$, which is analytically solvable only for some values of $\gamma$: $1/2$, $3/4$, $3/2$, $2$. After that one has to add the gravitational terms, obtain the super-momentum and super-Hamiltonian, and then solve them to write an expression in the form $\pi - h(H^G, d) = 0$ from the latter. In the cases of $\gamma = 3/2,\ 2$ the super-Hamiltonian is not analytically solvable, while in the case $\gamma = 3/4$, apart from the unphysical behaviour, the form of the $h$ function is so complicated that the quantization of the evolution operator is impossible. So one is forced to drop these models.\\
The only working case is $\gamma = 1/2$, which however is highly unphysical. It has a singular equation of state for a given energy density, while the velocity of sound diverges. However the BK mechanism works fine and gives, after discarding some unphysical solutions with the requirement for $h$ to be diff-invariant:
\begin{equation}\nonumber
\pi  = h(x, H^G, H^G_a)= \sqrt{\frac{qd}{(H^G)^2 - d}},
\end{equation}
where $d=H^G_aH^G_b q^{ab}$, which fulfils all the necessary conditions stated before. Hence an evolutionary picture can be recovered in this singular case. This suggests to investigate the possibility of \emph{ad hoc} modifications of the Lagrangian that could avoid this behaviour. In fact a similar case is analyzed  in \cite{thiemann:2006} by Thiemann who, by adding a constant term in the definition of $\Upsilon$, recovers an evolutionary picture and links the fluid to the k-essence, without singularities.

\section{The Schutz perfect fluid}
The model proposed by Schutz (see \cite{schutz:1970} \cite{schutz:1971}) for a baryonic fluid is based on the idea of expressing the motion of particles using six scalar fields over space-time: $\mu$, $\phi$, $\alpha$, $\beta$, $\theta$, $S$. Only the first and the last one have a physical meaning, one being the specific inertial mass, the other the entropy per baryon. They are combined in order to define the 4-velocity $
U_\nu = \mu^{-1}(\phi_{,\nu} + \alpha \beta_{,\nu} + \theta S_{,\nu}) = {v_\nu}{\mu}^{-1}$ and $\mu$ is then fixed by the normalization condition on it. The fluid action can be identified with the integral of the pressure of the fluid (see \cite{schutz:1970} \cite{schutz:1971}) and the calculation of the conjugate momenta follows the usual pattern. In this case one finds that only one momentum is independent: Schutz' is a constrained theory. In fact:
\begin{equation}\label{smomentum}
p_\phi =  \sqrt{q} \rho_0 \mu^{-1} v_\mu n^\mu = \pi,
\end{equation}
while $p_\alpha = p_\theta = 0$, $p_\beta = \alpha \pi$ and $p_S =\theta \pi$.
There are no secondary constraints, as one can check using the standard Dirac approach to constrained systems (see \cite{hanneaux:1994}). One can easily see that it is possible to write the Hamiltonian for the system only \emph{on-shell}, since otherwise it is impossible to express all the configuration variables in terms of the canonical ones.\\
By coupling the fluid with GR one expects the primary constraints to be simply the union of the two sets of primary constraints, since the fluid action does not contain any derivative of the metric tensor $g$. As for the matter-free theory the secondary constraints will be identified with the functionals that appear enforced by the lapse function and the shift vector in $\boldsymbol{H} = \int d^3 x \left(NH + N_a H^a \right)$,
which are, \emph{on-shell}:
\begin{subequations}\nonumber
\begin{align}
&H = \pm \sqrt{\frac{V}{\pi^2 - q \rho_0^2}}\left( \xi \pi^2 + \chi q \rho_0^2 \right)  + \sqrt{q} \rho_0 S T + H^G=0,\\
&H_a = \pi \phi_a + H^G_a=0,
\end{align}
\end{subequations}
where $\xi = (3, 1)$ and $\chi= \pm 1$.\\
Since the fluid action is added in a fully covariant way, and nothing could have damaged the diffeomorphism invariance of the model, one still expects these functionals to be the generators of the diffeomorphism group of space-time. So their algebra, calculated \emph{on-shell} in order to work with canonical variables only, is closed, and reproduces the standard Dirac algebra of vacuum GR. So one can also claim that there are no tertiary constraints since the Hamiltonian is a linear combination of such functionals.
At this point one is able to apply the BK mechanism.\\
Squaring the super-momentum and imposing it on the super-Ham\-iltonian, one can solve the latter for $\pi$, obtaining something in the form $\pi - h = 0$. Different choices for $\xi$ and $\chi$ lead to different results: one can select the physical one(s) among them requiring the new constraint to reduce to the form $\pi + \sqrt{q}\rho_0 = 0$ in the co-moving frame. This is the expected form from the definition of $\pi$ \eqref{smomentum}.\\
With this requirement only one case survives: $\xi = 1$ and $\chi= 1$. This gives:
\begin{equation}\nonumber
\pi = - \sqrt{q} \rho_0 \sqrt{\frac{  \left( 2d + \Xi^2 \pm \Xi \sqrt{8d + \Xi^2}  \right)}{2\left( d- \Xi^2 \right)}} = h,
\end{equation}
where $\Xi = \sqrt{q} \rho_0 S T + H^G$ and $d = H^G_a H^G_b q^{ab}$.\\
This is the candidate function to construct a physical Hamiltonian with.
It is a scalar density of weight one and it is the only allowed solution of the super-Hamiltonian constraint. So it fulfils all the conditions needed to be promoted to the role of generator of time evolution of observables. The striking result is that the entropy field $S$ enters directly in the time evolution operator, even if in a complicated way. Its role is not very clear, but the link between time and entropy is evident.\\
A much more interesting result is achieved in the co-moving frame, where $\Xi = 0$. Combining the expression of $\pi$ in that case with the expression of $\Xi$ stated above, one can write the identity $\pi = \frac{H^G}{ST}$, which, recalling the definition of the momentum conjugate to S, $p_S = \theta \pi$, becomes:
\begin{equation}\nonumber
S p_S = \frac{\theta H^G}{T} = \bar{h}.
\end{equation}
At this point one can identify the time variable $\tau$ with the logarithm of entropy, $\tau = ln S$: time evolution is directly linked to a naturally future pointing variable. 
This opens to applications in cosmology, while a natural development would be to implement the mechanism using the Ashtekar variables, getting closer to the formulation of Loop Quantum Gravity.


\end{document}